\documentclass[sigconf,screen]{acmart}

\usepackage{hyperref}
\usepackage{subcaption}
\usepackage{listings}

\AtBeginDocument{%
  \providecommand\BibTeX{{%
    \normalfont B\kern-0.5em{\scshape i\kern-0.25em b}\kern-0.8em\TeX}}}

\setcopyright{acmlicensed}
\copyrightyear{2018}
\acmYear{2018}
\acmDOI{XXXXXXX.XXXXXXX}

\acmConference[Conference acronym 'XX]{Make sure to enter the correct
  conference title from your rights confirmation email}{June 03--05,
  2018}{Woodstock, NY}
\acmISBN{978-1-4503-XXXX-X/18/06}

\begin{document}

\title{GenVectorX: A performance-portable SYCL library for Lorentz Vectors operations}

\author{Monica Dessole}
\orcid{0000-0002-2727-9123}
\affiliation{%
  \institution{CERN, EP SFT}
  \streetaddress{Esplanade des Particules, 1}
  \city{Geneva}
  \country{Switzerland}
  \postcode{1202}
}
\email{monica.dessole@cern.ch}

\author{Jolly Chen}
\affiliation{%
  \institution{CERN, EP SFT}
  \streetaddress{Esplanade des Particules, 1}
  \city{Geneva}
  \country{Switzerland}
  \postcode{1202}
}

\author{Axel Naumann}
\affiliation{%
  \institution{CERN, EP SFT}
  \streetaddress{Esplanade des Particules, 1}
  \city{Geneva}
  \country{Switzerland}
  \postcode{1202}
  }

\renewcommand{\shortauthors}{Dessole et al.}

\begin{abstract}
The Large Hadron Collider (LHC) at CERN will see an upgraded hardware configuration which will bring a new era of physics data taking and related computational challenges. To this end, it is necessary to exploit the ever increasing variety of computational architectures, featuring GPUs from multiple vendors and new accelerators. Performance portable frameworks, like SYCL, allow to offload the computational work on non-CPU resources, while retaining their performance, without the need to maintain different implementations of the same code. The High Energy Physics (HEP) community employs a wide variety of algorithms and tools for accelerators, but it still lacks a streamlined coherent approach that can target many use cases without compromising  the usability aspect.
In this paper, we present our efforts in creating GenVectorX, a C++ package that provides classes and functionalities to represent and manipulate particle events using the SYCL programming model. The SYCL-based implementation exhibits comparable performance and scalability as the CUDA implementation when targeting NVIDIA GPUs.
\end{abstract}



\received{20 February 2007}
\received[revised]{12 March 2009}
\received[accepted]{5 June 2009}

\maketitle
\sloppy

\section{Introduction}
The core of High Energy Physics (HEP) research is characterised by the need for processing and analysing huge amounts of data coming from the accelerators. The largest source of such data is the Large Hadron Collider (LHC), hosted at CERN in Switzerland, which since its start has reached peaks of 1 PB/s of data generated from physics events that need to be stored and accessed by the physics community to be analysed. To this extent, the Worldwide LHC Computing Grid (WLCG) \cite{annurev-nucl-102010-130059} was developed in cooperation between CERN and other research institutes as a shared computing infrastructure serving all interested scientists around the world. Alongside this main distributed facility, it is common to have smaller computing clusters at the level of the single research institution. All these computing facilities exhibit different hardware configurations and composition. The demand for compute power and data storage is expected to increase dramatically in the next years, as in 2029 the High Luminosity LHC (HL-LHC), an upgraded hardware configuration of the LHC particle accelerator, will start operating with a demand of computational resources that is estimated between 50 and 100 times with respect to the current use \cite{hep2019roadmap}.
In this scenario, in which scientists from all over the world need to deal with an ever increasing amount of input data and to run their analysis on heterogeneous platforms, performance portability plays a crucial role. Such considerations motivate the need for developing performance portable software tailored to the HEP use case. Although each research group may need to tweak their analysis to use specific libraries, most software environments in HEP use ROOT \cite{BRUN199781} as the tool for storing, analyzing and visualizing physics data regarding particle collisions. These collision events are expressed as operations on particles, represented as $4$-dimensional time-space vectors, also known as Lorentz Vectors. Within ROOT, the GenVector package contains classes for specialized vectors in  $2$, $3$, and, in particular, $4$ dimensions, and their operations, providing models and capabilities tailored to HEP analysis.

Performance portability of scientific computing applications is gaining importance as hardware becomes more heterogeneous. While NVIDIA GPUs are now standard co-processors in generic purpose computations, other vendors' GPU architectures are emerging in the High Performance Computing (HPC) scenario, see e.g. LUMI HPC system which is based on AMD GPUs and Aurora HPC system on Intel(R) GPUs. Moreover, the landscape of multi-core CPUs and alternative accelerators is also diversifying, see e.g. FPGAs and the RISC-V initiative \cite{3597047}. Programming techniques for GPUs and other accelerators are very different from that of traditional CPUs, requiring a significant understanding of the hardware architecture to achieve good performance. Special languages and compilers have been developed to target these architectures, such as CUDA \cite{cuda} for NVIDIA GPUs and HIP \cite{hip} for AMD GPUs. Porting scientific codes to new hardware is not sustainable as it costs valuable developer time due to the adoption of hardware-specific programming languages. Performance portability frameworks overcome this issue by introducing an abstraction layer that can map Single Instruction Multiple Data (SIMD) parallel models across different hardware platforms. This concept has been successfully demonstrated, for example, by libraries like Kokkos \cite{CARTEREDWARDS20143202}, Alpaka \cite{ZenkerAsHES2016}, as well as SYCL \cite{sycl}, providing abstractions to enable heterogeneous device programming.
In this work, we focus on SYCL because of its performance efficiency and portability on different devices, such as CPUs, GPUs, and FPGAs. SYCL is a cross-platform abstraction layer inspired by OpenCL that enables the composition of a "single-source" code using modern ISO C++, that can be compiled for heterogeneous architectures. SYCL’s increasing popularity has led to the development of a diverse range of implementations and compilers within its ecosystem. Among them, the most notable are Intel(R) oneAPI Toolkit \cite{oneapi} and AdaptiveCpp \cite{acpp}.

In this paper we extend GenVector to GenVectorX, an accelerated library that provides both a CUDA and a SYCL implementation of the Lorentz Vector classes that facilitate computations with physical vectors. Our analysis focuses on the impact of manual code specialization upon developers with regards to code maintenance, with the explicit scope of minimizing code duplication and maximizing code reuse in order to promote code sustainability and portability. We compare the performance of some of the most common operations involving Lorentz Vectors on multiple platforms. We carry out an extensive test campaign on NVIDIA GPUs, with particular focus on the performance gap between native CUDA and SYCL code execution. Focusing on an invariant mass computation problem, we study scaling and demonstrate that reach performance portability, for almost all sizes of inputs. 

This paper is organized as follows. Section \ref{sec:related_work} briefly reviews previous performance portability efforts, with focus on the HEP community. In Section \ref{sec:background}, we provide a detailed background on the organization of the GenVector library and the SYCL execution model. Section \ref{sec:porting} details the porting procedure and provides quantitative results about code divergence. Performance analysis results are presented in Section \ref{sec:performance} and Section \ref{sec:conclusion} concludes this treatise.

Specifically, the contributions of this paper are as follows:
\begin{itemize}
    \item We detail the migration of a large, complex, C++ code base to both SYCL and CUDA, providing guidance and insights regarding the analogies and differences between the two frameworks for other developers interested in migrating their own codes.
    \item We provide a detailed performance analysis of the migrated SYCL code on different platforms and architectures.
    \item  We compute and compare the performance portability and code divergence of GenVectorX, to explore the trade-offs between maintaining a single source code and specializing small regions of code for specific targets.
\end{itemize}

\section{Related Work} \label{sec:related_work}

The High Energy Physics community employs a wide variety of algorithms and tools for accelerators, but it still lacks a streamlined coherent approach that can target many use cases without compromising the usability aspect, although there have been preliminary, see \cite{Bocci_2023,atif2023evaluating,bhattacharya2022portability}. However, to the best of our knowledge, none of them has been focusing on the SYCL framework. Moreover, this work is dedicated to the porting of a fundamental library, whose benefits could reflect on a wide variety of related software. 

A number of studies have detailed the process of migrating from CUDA to SYCL, see \cite{PPH23, RPP23, 10024604, ChristgauSteinke2020,0780269,3585372}. This paper provides insights about developing CUDA and SYCL extensions of codes that are still under active development, thus with the explicit scope of minimizing code duplication and maximizing code reuse in order to promote sustainability and together with portability. Additionally, by investigating the performance of the migrated code on modern platforms, we are able to identify potential improvements to the migration process that could improve the experience for future developers.

\section{Background} \label{sec:background}

\subsection{The GenVector Library}

ROOT \cite{BRUN199781} is a library for storing, analyzing and visualizing physics data regarding particle collisions used by particle physicists spread all over the world. Within ROOT, the GenVector is a large ($\sim$ 11k lines of code) package is intended to provide classes and functionalities to represent physicals vectors and transformations (such as rotations and Lorentz transformations) in $2$, $3$, and $4$ dimensions, according to the needs of High Energy Physics researchers.
Here we briefly describe some of the classes, along with all the member functions and associated functions which users can count on.

There is a user-controlled freedom as to how the vector is internally represented. This is expressed by the choice of a coordinate system, which is supplied as a template parameter when the vector is constructed. A coordinate system can be one of several choices (Cartesian, Polar, and so forth) in $2$, $3$, or $4$ dimensions, and the choice of the coordinate system simultaneously determines the dimension of the vector. There is a further degree of control: each coordinate system is itself a template so that the user can specify the underlying scalar type, say single or double precision floating point numbers. For example, \texttt{Cartesian3D<double>} is a coordinate system representing $3$-dimensional vectors in Cartesian coordinates.
Transformations are modeled as non-template classes, using double as scalar-type. For the purposes of understanding the classes available, the transformations are grouped as follows: Rotations (in two and three dimensions), Lorentz transformations, and Poincare transformations, which are Translation/Rotation combinations. Each group has several members, which may model physically equivalent transformations but with different internal representations. For example, a Rotation may be expressed as a $3\times3$ matrix (\texttt{Rotation3D}) or as an axis and angle of rotation (\texttt{AxisAngle}). Some specializations of the templates are given special names for convenience and brevity. For example, \texttt{XYZVector} is typedefed to \texttt{DisplacementVector< Cartesian3D<double> >}.

Every class -- we include every class obtained by instantiating some template -- is in the namespace \texttt{ROOT::Experimental::}. Every class is default constructible, copy constructible, and copy assignable. Every class is dependent on some real type to be used for scalars. Every class is equality comparable (for equals and not-equals) against objects of the identical type. 
Every class has a pair of corresponding headers: a header containing the full class or class template definition, matching the name of the class (for example, \texttt{Rotation3D.h}) and a “forwarding header” with the letters fwd, containing only declarations of the class or class template (for example, \texttt{Rotation3Dfwd.h}). Headers freely include each other’s forwarding headers, but inclusion of the full headers is organized so as to minimize coupling and avoid circular dependencies.

\subsection{SYCL Computational Model}
SYCL is a Khronos Group language standard that enables code for heterogeneous and offload processors, to be written using modern ISO C++, providing APIs and abstractions for finding devices (e.g. CPUs, GPUs, FPGAs) on which code can be executed, and to manage data resources and code execution on those devices. The SYCL kernel consists of the main computational kernel, which can be expressed as a C++ lambda function or as a functor object, the argument values associated with the kernel, and the parameters defining index range. Similarly to the CUDA execution model, the SYCL index hierarchy also consists of a $1$, $2$, or $3$-dimensional grid of work-items, corresponding to the single execution threads. These work-items are grouped into equal sized thread groups called work-groups. Threads in a work-group are further divided into equal sized vector groups called sub-groups. Within a sub-group, it is possible to perform collective operations such as broadcast, shuffle, reduction, and so on. Beside the main global memory, SYCL defines a local memory which is shared among all the work-items within a work-group and which can be exploited to enhance data reuse. Depending on the implementation and the hardware availability, this shared local memory can be mapped into different physical memories. Furthermore, the work-items within a work-group can be synchronized with the use of barriers, but synchronization across different work-groups is not possible.

The SYCL standard defines two abstractions for declaring and accessing data on devices with different memory contexts: buffers and Unified Shared Memory (USM) pointers. With buffers, the data management is handled entirely by the SYCL runtime and access to the underlying data is permitted via accessors. When accessing a buffer on either the host or in a device kernel, the user needs to create an accessor that defines the type of access (read-only, write-only, or read-write), which is information that the runtime uses to determine the data dependencies and necessary memory transfers. For USM pointers, there are three different allocation types: host, device, and shared. These pointers allow for direct dereferencing and fine-grained control over the ownership, but the data dependencies and copy operations need to be defined explicitly by the user. In this paper, we experiment with buffers where the data can live on both the host and device, and with device pointers where the data is allocated and directly accessible on the device only.

In SYCL, kernel functions are enqueued on a device queue, i.e. a kernel scheduler on any SYCL device, in order to be executed on a particular device. The return type of the SYCL kernel function is void, and all memory accesses between host and device are through accessors or through USM pointers. There are two ways of defining kernels: as named function objects or as lambda functions. A backend may also provide interoperability interfaces for defining kernels. When function objects are used, these provide the same functionality as any C++ function object, with the restriction that they need to follow SYCL rules to be device copyable. The kernel function can be templated via templating the kernel function object type. The operator() member function must be const-qualified, and it may take different parameters depending on the data accesses defined for the specific kernel. 
Kernels may also be defined as lambda functions. The name of a lambda function in SYCL may optionally be specified by passing it as a template parameter to the invoking member function, and in that case, the lambda name is a C++ typename which must be forward declarable at namespace scope. If the lambda function relies on template arguments, then if specified, the name of the lambda function must contain those template arguments which must also be forward declarable at namespace scope. The class used for the name of a lambda function is only used for naming purposes and is not required to be defined.

\section{Migration Process} \label{sec:porting}

\subsection{SYCL and CUDA extentions}
As presented in the previous Section, GenVector is a C++ package intended to provide classes and functionalities to represent physicals vectors and transformations in multiple physical dimensions, meaning that it mainly provides data structures for HEP analysis.

With C++ being the programming language of choice, almost all classes and methods can be ported to SYCL and CUDA in a straightforward manner by providing macros and wrapper functions that encapsulate the differences between the host code for the two programming models. 
The main necessary changes involve mathematical functions. In fact, in SYCL the OpenCL math functions are available in the namespace \texttt{sycl::} on host and device with the same precision guarantees as defined in the OpenCL 1.2 specification document \cite{opencl} for host and device. For a SYCL platform the numerical requirements for host need to match the numerical requirements of the OpenCL math built-in functions. In GenVectorX, all  basic mathematical functions belong to the namespace \texttt{ROOT::Experimental::} and they are are compiled and guarded according to a macro definition.
Thus, when SYCL is enabled, the mathematical functions corresponds to their SYCL equivalent. Similarly, when CUDA is enabled, all mathematical functions are defined according to their CUDA equivalent. Unlike SYCL, where classes and methods does not need to be decorated to be called on the device, CUDA requires the use of both \texttt{\_\_host\_\_} and \texttt{\_\_device\_\_} decorators, indicating that the classes and methods can be called by both the host and the device. In order to maintain a single source for CUDA, SYCL and standard CPU, we use instead decorators defined as macros which are defined as blank whenever CUDA is disabled.

Some methods make an exception and cannot be ported to SYCL. 
In particular, every class in GenVector has methods meant for exception handling which cannot be ported as is. Also, we find operators \texttt{<<} and \texttt{>>} to ostreams and from istreams respectively, which produce and read back human-readable output representing the vectors or transformations. However, the functionalities related to I/O operations are compromised, as they usually assume sequentiality of the execution, while, for example, the access to the standard output is asynchronous in massively parallel architectures. For this reason these methods are dropped.

The strategies described so far allow to have a single source code defining classes and functionalities to manipulate physicals vectors that can be used in code running on both host and device. Such object can be included in used defined code targeting a CUDA or SYCL device. In order to provide user-ready functionalities, we define higher level kernel functions to carry our most common computations in High energy physics analysis. Such functions are defined as \texttt{\_\_global\_\_} CUDA kernels and, for what concerns SYCL, as function objects. As an example, we detail two function kernels that with be used later on the performance analysis, namely the Invariant Masses and Boosting computation. The \texttt{InvariantMasses} function returns the invariant mass of two particles expressed in any $4$-dimensional coordinate system. This function is defined accordingly to the targeted platform, handling the device memory allocation and transfers if necessary. Figures \ref{fig:CUDAkernel} and \ref{fig:SYCLkernel} show the CUDA kernel and the SYCL object function definitions, respectively. It is worth noticing that both kernels can handle a template Lorentz Vector object that exposes mass computation via the method \texttt{m()}. The function \texttt{ApplyBoost} applies a $4$-dimensional Lorentz transformation represented internally by a $4 \times 4$ orthosymplectic matrix. Its structure is similar to that the \texttt{InvariantMasses} function. 

\begin{figure}
    \centering
\begin{verbatim}
template <class Scalar, class LVector>
__global__ void InvariantMassesKernel
(LVector *v1, LVector *v2, Scalar *m, size_t N)
{
    int id = blockDim.x * blockIdx.x + threadIdx.x;
    if (id < N)
    {
        LVector w = v1[id] + v2[id];
        m[id] = w.mass();
    }
}
\end{verbatim}    
\caption{Invariant Masses CUDA Kernel}
    \label{fig:CUDAkernel}
\end{figure}

\begin{figure}
    \centering
\begin{verbatim}
template <class Scalar, class Vector>
class InvariantMassesKernel
{
public:
    InvariantMassesKernel
    (LVector *v1, LVector *v2, Scalar *m, size_t n)
        : d_v1(v1), d_v2(v2), d_m(m), N(n) {}

    void operator()(sycl::nd_item<1> item) const
    {
        size_t id = item.get_global_id().get(0);
        if (id < N)
        {
            LVector w = d_v1[id] + d_v2[id];
            d_m[id] = w.mass();
        }
    }

private:
    LVector d_v1;
    LVector d_v2;
    Scalar d_m;
    size_t N;
};
\end{verbatim}    
\caption{Invariant Masses SYCL function object}
    \label{fig:SYCLkernel}
\end{figure}

\subsection{Code Divergence}
In this paper, our analysis focuses on the impact of manual code specialization upon developers with regards to code maintenance. 
We measure this impact using a version of the code divergence metric \cite{8639933, PENNYCOOK2019947} which represents the average pairwise distance between source codes used to target different platforms. For the sake of clarity, let us formally introduce the following terminology: a \textit{problem} is an input to application, with a correctness test and observable performance; an \textit{application} is a collection of software that can run problems on one or more platforms; a \textit{platform} is a collection of hardware and software on which an application runs problems. 
Code divergence is calculated as:
\begin{equation}
    CD(a,p,H) = \binom{|H|}{2}^{-1} \sum_{(i,j) \in H \times H} d_{i,j}(p,a)  \label{eq:code_divergence}
\end{equation}
where $d_{i,j}(p,a)$ represents the distance between the source code required to solve problem $p$ using application $a$ on platforms $i$ and $j$ (from platform set $H$).
As done in \cite{RPP23, 9484790}, we adopt the Jaccard distance defined as $d_{i,j}(p,a) = 1-s_{i,j}(p,a)$, where $s_{i,j}(p,a)$ represents the similarity between two source codes:
\begin{equation}
    s_{i,j}(p,a) = \left| \frac{c_i(a,p) \cap c_j(a,p)}{c_i(a,p) \cup c_j(a,p)} \right|. \label{eq:code_similarity}
\end{equation}
Here, $c_i$ and $c_j$ represent the set of source lines required to compile application $a$ and execute problem $p$ for platforms $i$ and $j$, respectively. Note that the similarity metric takes values in the interval $[0, 1]$, where $1$ indicates that all code is shared within the platforms and $0$ means that the entire code is specialized for each platform.

We focus on GenvectorX as target application and state in Table \ref{tab:code_divergence} the values of code similatiry of different problems between CPU execution, i.e. the GenVector original code, and the other platforms taken into account. Since SYCL has a single source code regardless the hardware targeted, here it is considered as a single platform. Since code divergence and convergence are normalized to the size
of the code base, we also present a breakdown of raw source lines of code for the numerator and denominator, denominated as union and intersection respectively, of the code similarity \eqref{eq:code_similarity}, excluding whitespaces and comments. We can denote that the SYCL and CUDA variants share almost all the code with the CPU implementation. The total code divergence evaluated according to equation \eqref{eq:code_divergence} is $0.079$.

\begin{table}[ht]
  \caption{Code Similarity of GenvectorX vs GenVector}
  \label{tab:code_divergence}
  \begin{tabular}{ccccc}
    \toprule
    Similarity & Union & Intersection & Platform & Problem   \\
    \midrule
    0.9694 & 10365 & 10048 & CUDA & Invariant Masses\\
    0.9715 & 10276 & 9983 & SYCL & Invariant Masses\\
    0.9693 & 10358 & 10041 & CUDA & Boosting\\
    0.9716 & 10335 & 10042 &SYCL & Boosting\\
  \bottomrule
\end{tabular}
\end{table}

The computation of performance portability metric \cite{8639933} requires the efficiency $e_{i}(p,a)$, i.e. the efficiency of problem $p$ solved within application $a$ on platform $i$. For example, the efficiency can be expressed as the ratio of the actual execution time and the best execution time known according to some analytical model, so that it takes values in the range $[0,1]$. The computation and tuning of an analytical performance modal can be far from trivial and it is outside the scope of the current work.

\section{Performance Analysis} \label{sec:performance}

\subsection{Experimental Setup}
We compare the GenVectorX library on three different computing environment:
\begin{enumerate}
    \item Intel(R) Xeon(R) Platinum 8362 CPU @ 2.80GHz with NVIDIA A100 40GB PCIe using CUDA 12.2 \label{olgpu}
    \item AMD Ryzen 7 5700G with Radeon Graphics with  NVIDIA GeForce RTX 3060 using CUDA 12.2 \label{victus}
    \item Intel(R) Xeon(R) Gold 6336Y CPU @ 2.40GHz with NVIDIA L4 using CUDA 12.3 \label{olice}
\end{enumerate}
The main features of the GPUs that are relevant in our case are summarized in Table \ref{tab:gpu_specs}, while relevant information about CPUs are reported in Table \ref{tab:cpu_specs}.
\begin{table}[ht]
  \caption{GPUs specification}
  \label{tab:gpu_specs}
  \begin{tabular}{lccc}
    \toprule
     & A100 & RTX 3060 & L4   \\
    \midrule
    CUDA Capability & 80 & 86 & 89\\
    Global Memory (GB) & 40 & 12 & 24 \\
    FP32 Peak (TFLOPs) & 19.5 & 12.74 & 30.3\\
    Bandwidth (GB/s) & 1555.0 & 360.0 & 300.0 \\
  \bottomrule
\end{tabular}
\end{table}

\begin{table}[ht]
  \caption{CPUs specification}
  \label{tab:cpu_specs}
  \begin{tabular}{lccc}
    \toprule
     & Cores & Threads & Peak (GHz)  \\
    \midrule
    Xeon(R) Platinum 8362 & 64 & 1 & 3.6\\
    Ryzen 7 5700G & 16 & 2 & 3.8 \\
    Xeon(R) Gold 6336Y & 96 & 2 & 3.6 \\
  \bottomrule
\end{tabular}
\end{table}

We compare two different implementations of SYCL compilers, i.e. Intel(R) oneAPI Toolkit \cite{oneapi} and AdaptiveCpp \cite{acpp}. We investigate the performance of GenVectorX on two main problems, namely the computation of invariant masses from two arrays of particles and the boosting of one array of particles. 

We collect results using both buffers and device pointers to manage the memory in SYCL and using both Intel(R) OneAPI and AdaptiveCpp as compilers. All results shown are compared against a single threaded CPU computation. Indeed, the scope of the current work is to collect results concerning the performance portability efficiency of SYCL, rather then proving e.g. that we can benefit from the use of GPUs in this context.

\subsection{Scaling of SYCL vs CUDA}
\begin{figure}
    \centering
    \includegraphics[width=.7\linewidth]{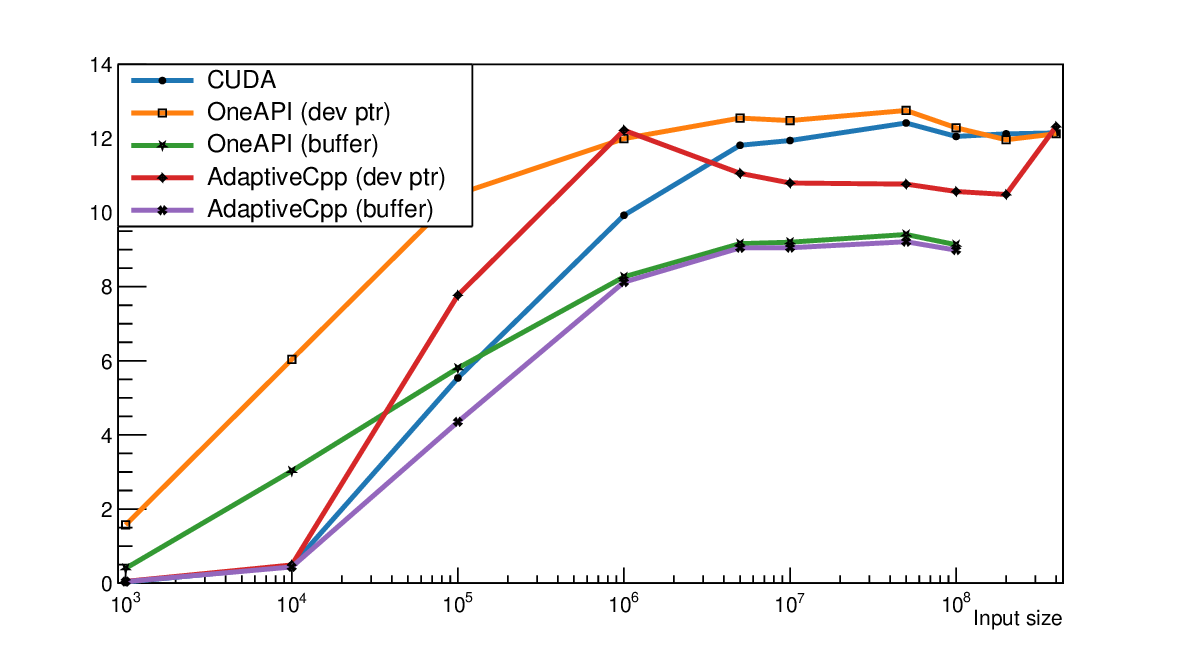}
    \includegraphics[width=.7\linewidth]{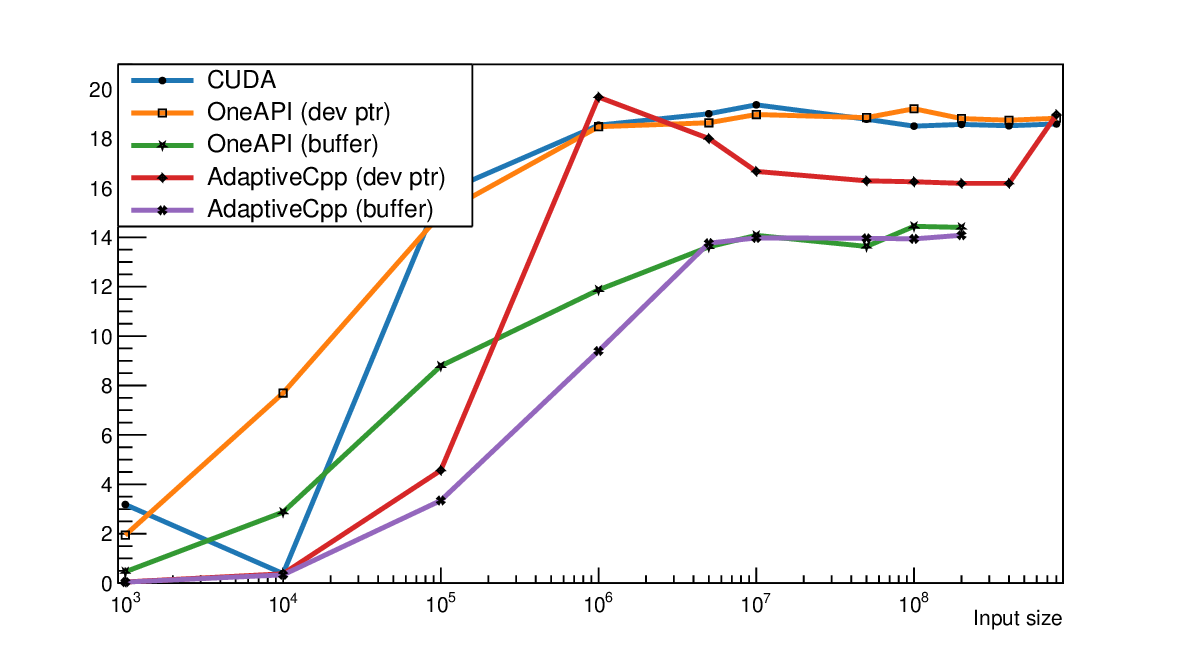} 
    \caption{Weak scaling against single threaded execution on computing environment \ref{olgpu} for the Boosting test on CUDA backend. Double precision results are shown on the top, while single precision ones are shown on the bottom.}
    \label{fig:boost_olgpu_cuda}
\end{figure}

\begin{figure}
    \centering
    \includegraphics[width=.7\linewidth]{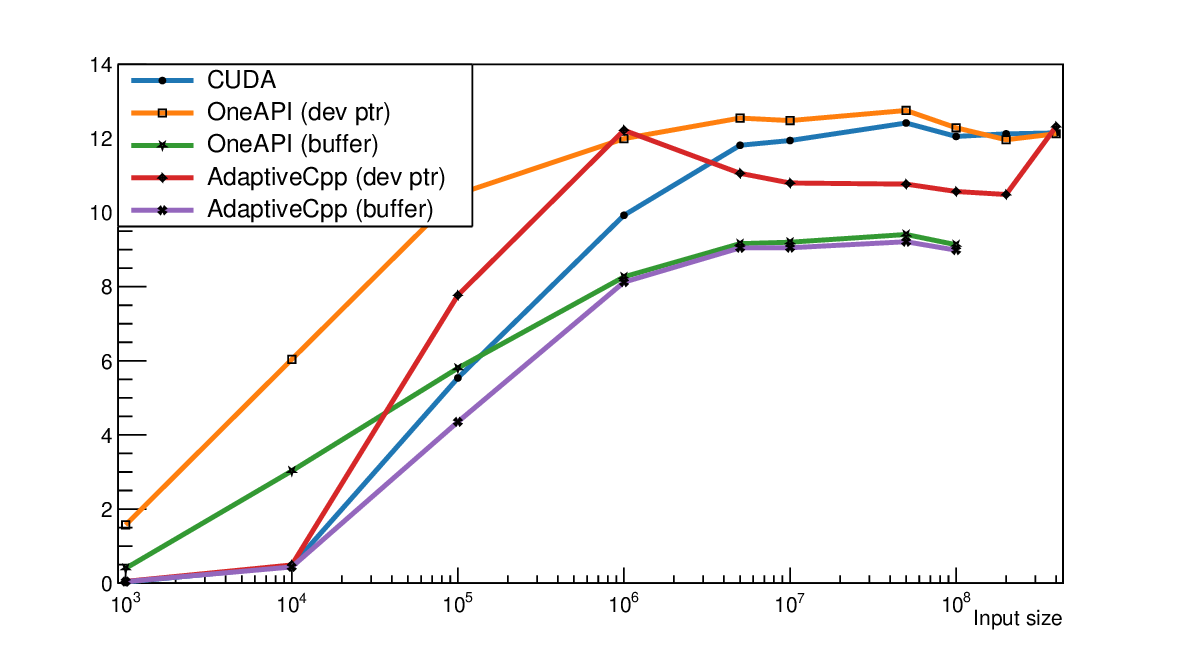}
    \includegraphics[width=.7\linewidth]{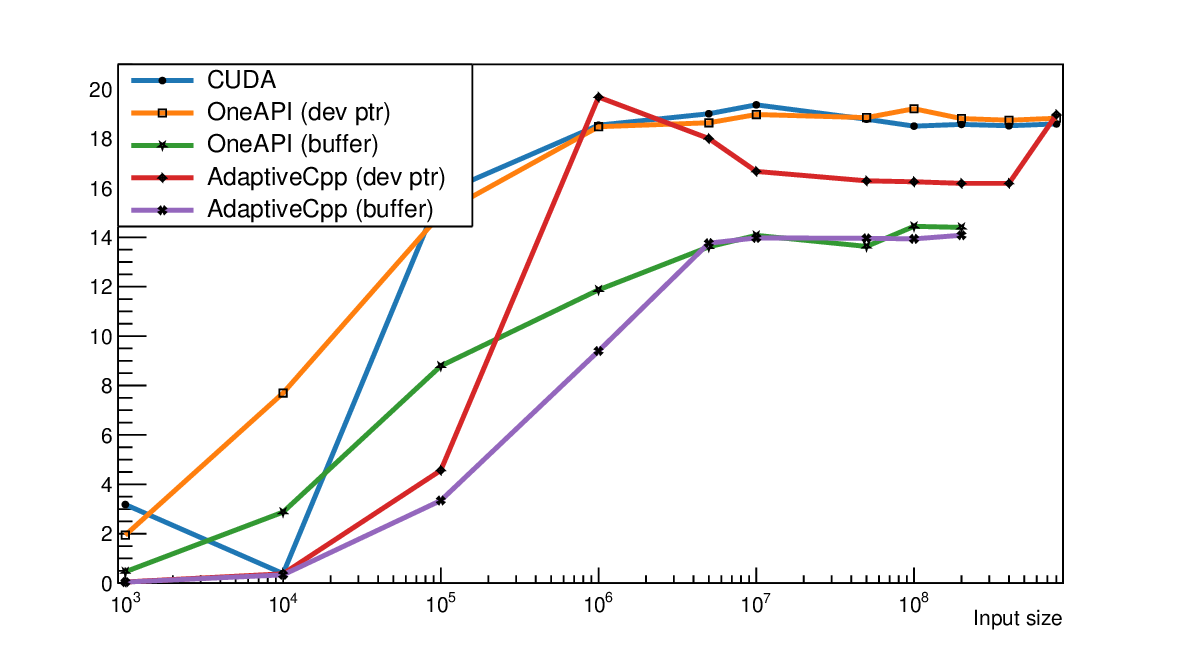} 
    \caption{Weak scaling against single threaded execution on computing environment \ref{olgpu} for the Invariant Masses test on CUDA backend. Double precision results are shown on the top, while single precision ones are shown on the bottom.}
    \label{fig:invms_olgpu_cuda}
\end{figure}

\begin{figure}
    \centering
    \includegraphics[width=.7\linewidth]{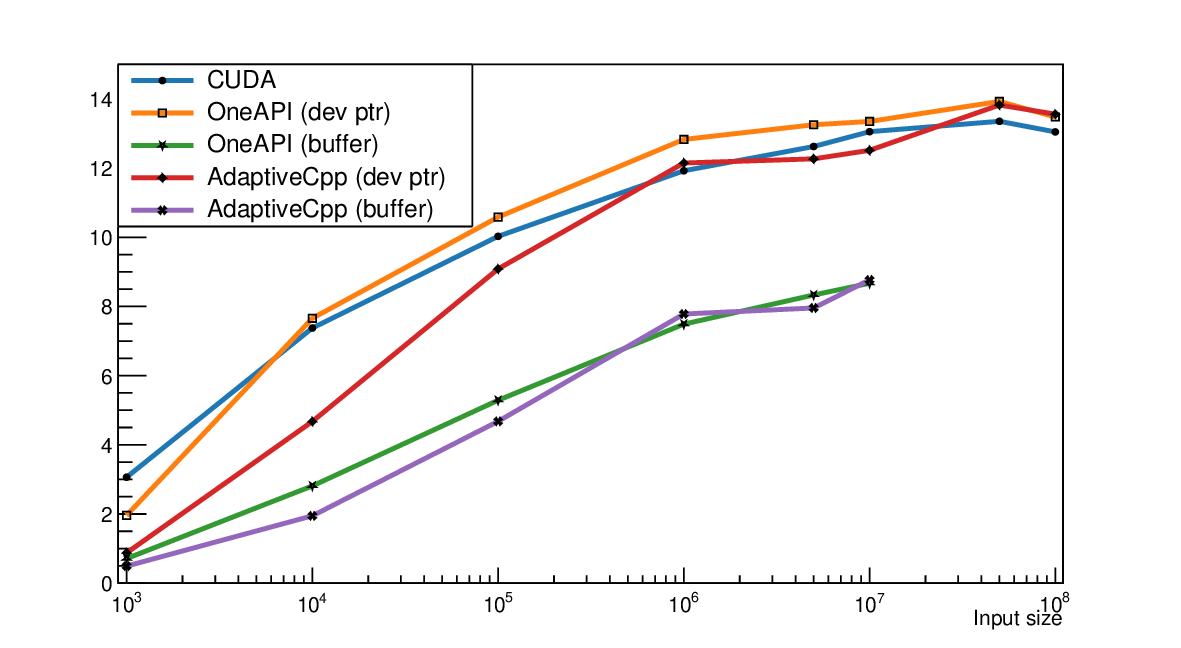}
    \includegraphics[width=.7\linewidth]{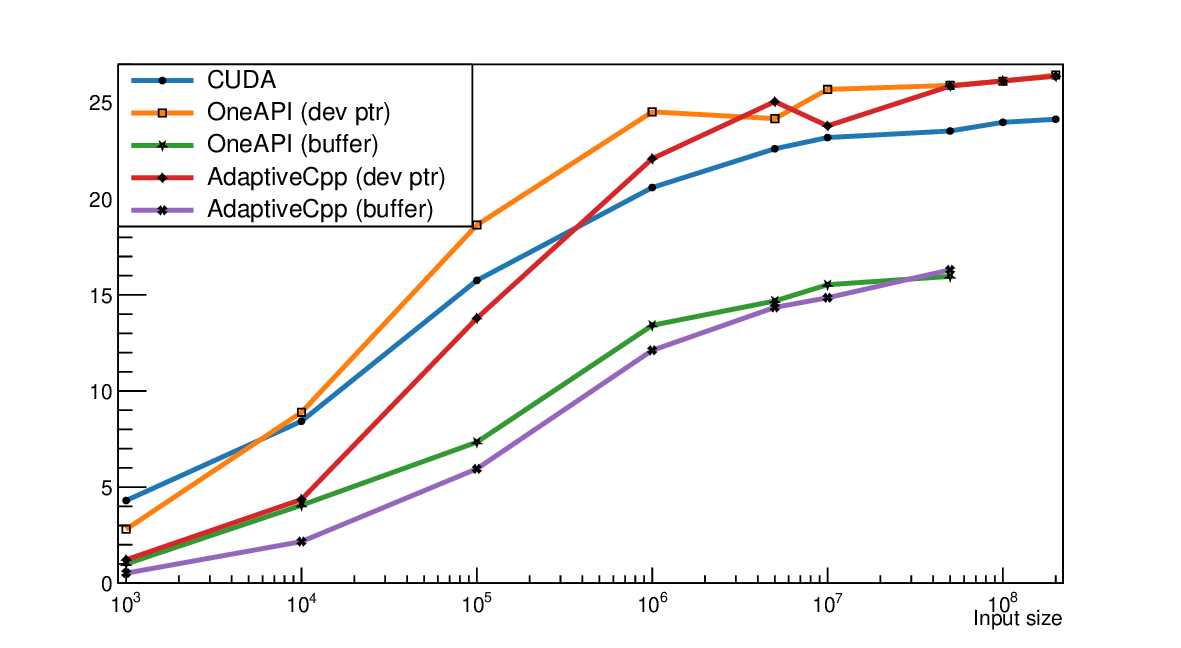}

    \caption{Weak scaling against single threaded execution on computing environment \ref{victus} for the Boosting test on CUDA backend. Double precision results are shown on the top, while single precision ones are shown on the bottom.}
    \label{fig:boost_victus_cuda}
\end{figure}

\begin{figure}
    \centering
    \includegraphics[width=.7\linewidth]{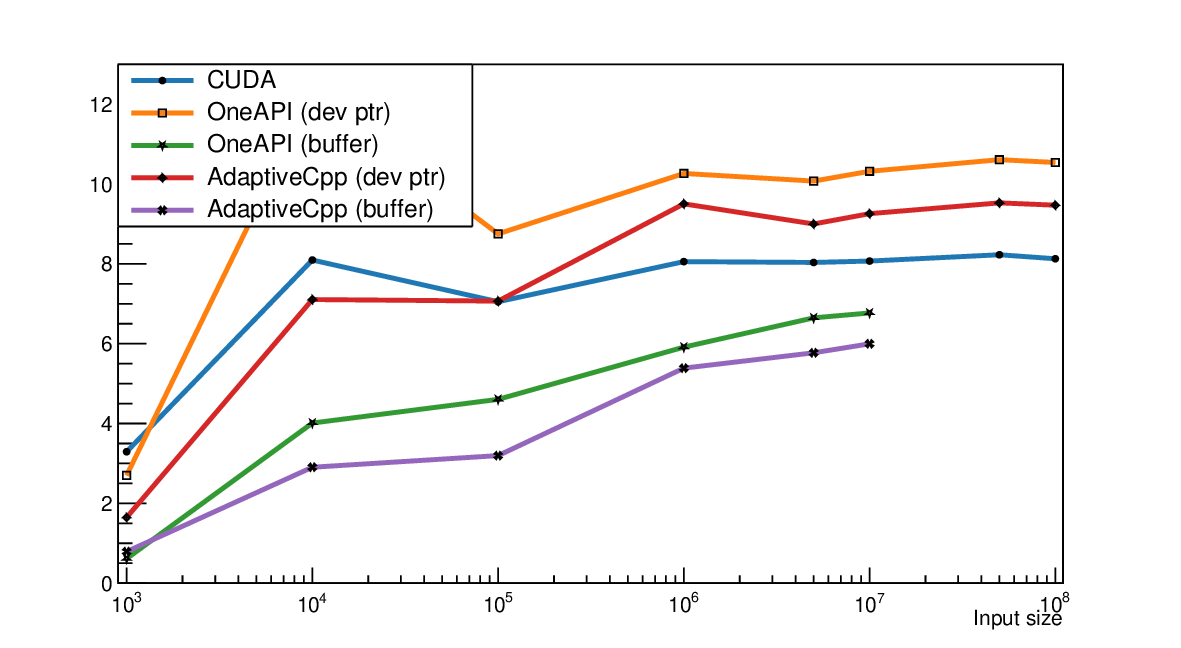}
    \includegraphics[width=.7\linewidth]{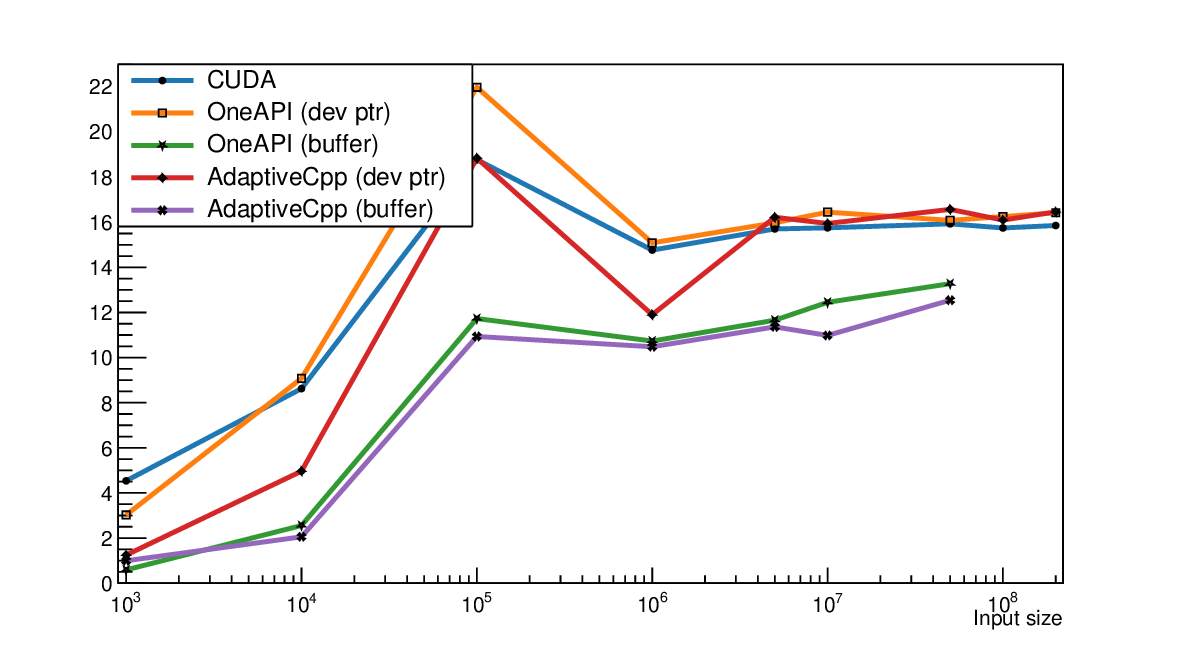} 

    \caption{Weak scaling against single threaded execution on computing environment \ref{victus} for the Invariant Masses test on CUDA backend. Double precision results are shown on the top, while single precision ones are shown on the bottom.}
    \label{fig:invms_victus_cuda}
\end{figure}

\begin{figure}
    \centering
    \includegraphics[width=.7\linewidth]{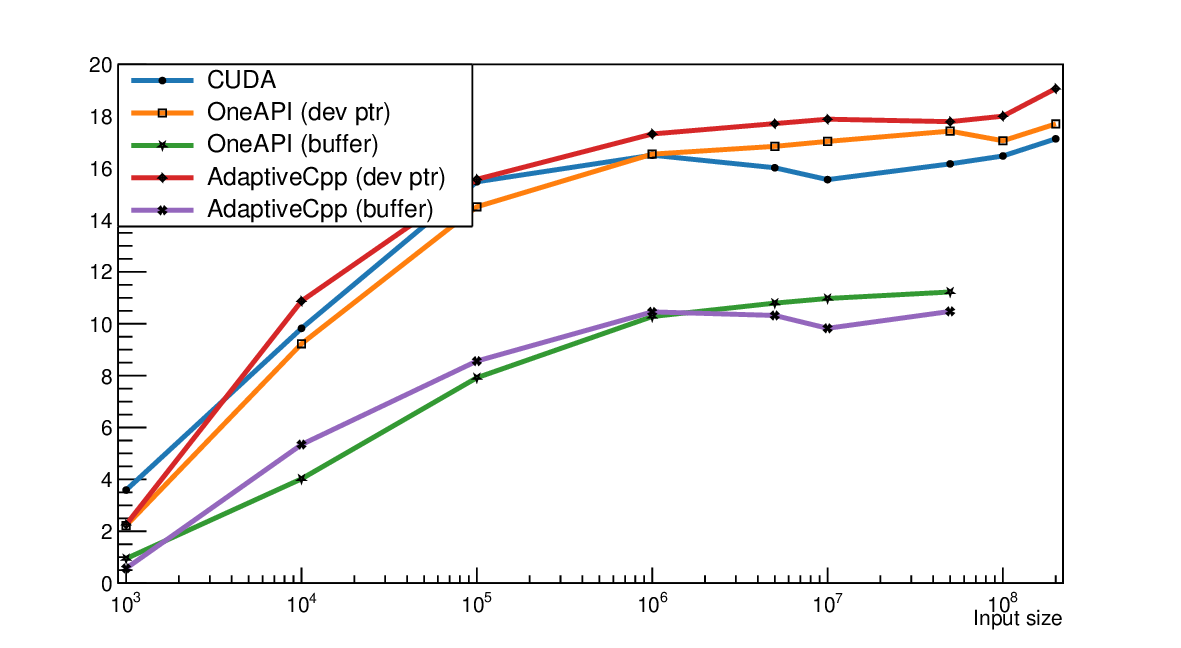}
    \includegraphics[width=.7\linewidth]{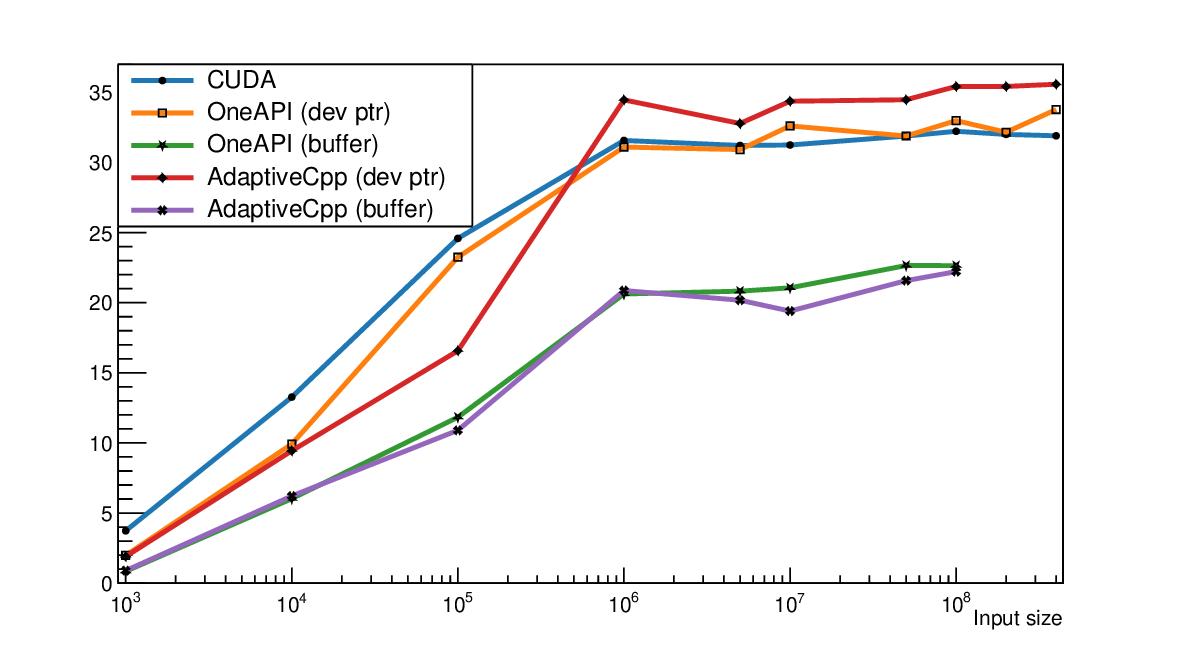}

    \caption{Weak scaling against single threaded execution on computing environment \ref{olice} for the Boosting test on CUDA backend. Double precision results are shown on the top, while single precision ones are shown on the bottom.}
    \label{fig:boost_olice_cuda}
\end{figure}

\begin{figure}
    \centering
    
    \includegraphics[width=.7\linewidth]{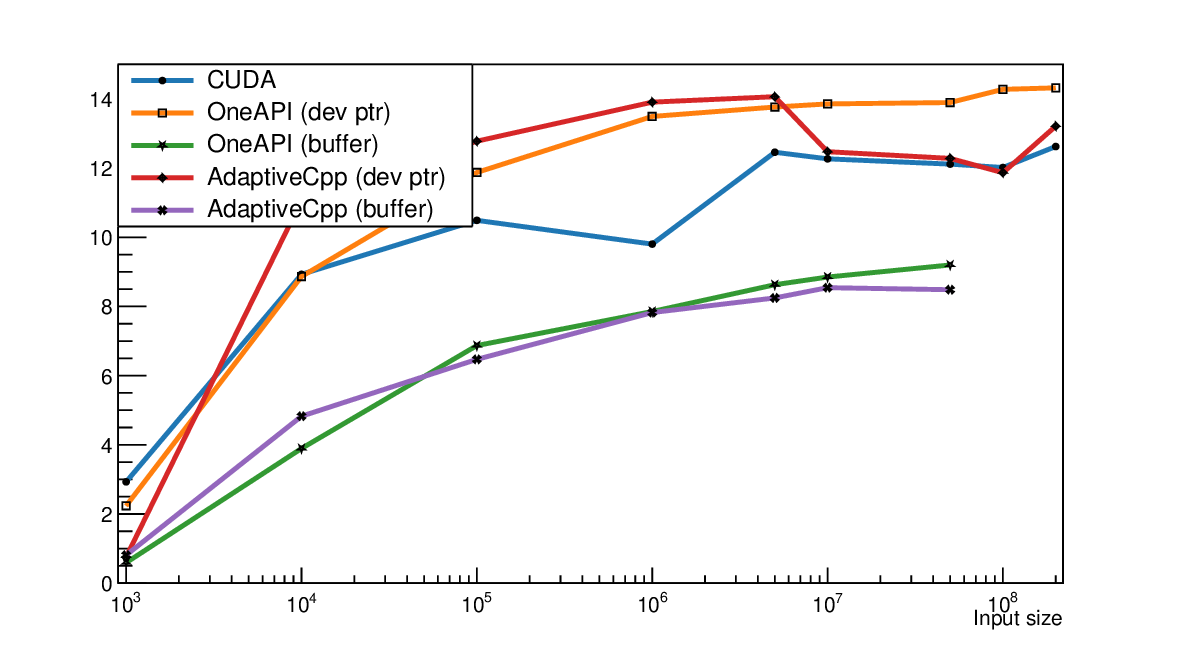}
    \includegraphics[width=.7\linewidth]{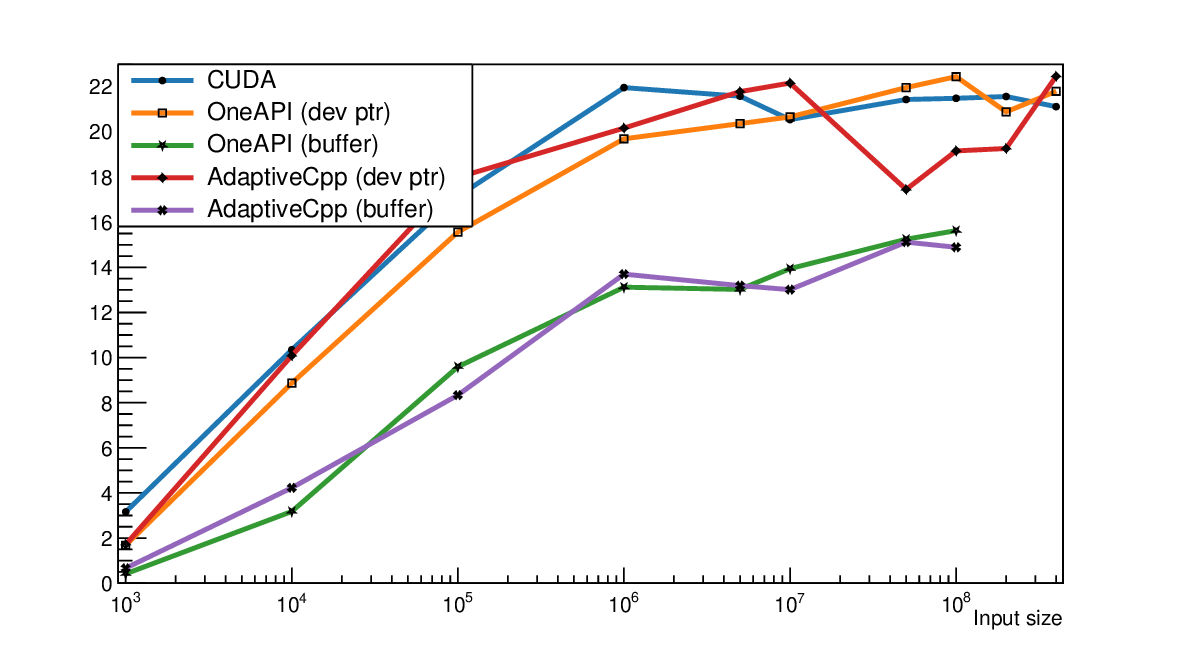} 

    \caption{Weak scaling against single threaded execution on computing environment \ref{olice} for the Invariant Masses test on CUDA backend. Double precision results are shown on the top, while single precision ones are shown on the bottom.}
    \label{fig:invms_olice_cuda}
\end{figure}

We evaluate the performance portability of SYCL by evaluating its weak scaling and comparing it against that of CUDA. Figures \ref{fig:boost_olgpu_cuda} and \ref{fig:invms_olgpu_cuda} show results on computing environment \ref{olgpu}, Figures \ref{fig:boost_victus_cuda} and \ref{fig:invms_victus_cuda} on \ref{victus} and Figures \ref{fig:boost_olice_cuda} and \ref{fig:invms_olice_cuda} on \ref{olice}, respectively. Figures \ref{fig:boost_olgpu_cuda}, \ref{fig:boost_victus_cuda}, \ref{fig:boost_olice_cuda} show results corresponding to the Boosting test and Figures \ref{fig:invms_olgpu_cuda}, \ref{fig:invms_victus_cuda}, \ref{fig:invms_olice_cuda} show results for the Invariant Masses test.
Each figure is composed by two subfigures, in which subfigures on the left refer to double precision computation, while subfigures on the right to single precision. Each subfigure shows the scaling obtained with native CUDA, with Intel(R) OneAPI compiler with the use of device pointers and buffers, as well as with AdaptiveCpp compiler with the use of device pointers and buffers. Scaling is shown in function of the input sizes and it is computed as the ratio between the timing of a one threaded CPU execution and the target execution. For every measurement, the timing is evaluated as the best execution time out of three runs executed sequentially, on the same queue for SYCL and on the same stream for CUDA. Results highlight that when buffers are used to manage memory transfers and access, CUDA outperforms SYCL, whatever compiler is adopted. On the other hand, when device pointers are used instead of buffers, Intel(R) oneAPI and AdaptiveCpp perform similarly to CUDA and even outperform it. It is interesting to point out that Figures \ref{fig:boost_olgpu_cuda} --  \ref{fig:invms_olice_cuda} don't show results for SYCL with buffers with either Intel OneAPI or AdaptiveCpp when large input sizes are involved. Indeed, SYCL fails to manage the memory with buffers and accessors on CUDA devices, giving a CUDAMalloc failure. 
In SYCL, one way to synchronize memory is by destructing the buffer after computation is complete. In our code, this is accomplished by having a scope around the buffer creation and kernel function call. As this issue does not affect the execution with SYCL device pointers, it is nor clear what can be the source code of this difference in the behaviour.

\subsection{Weak Scaling of SYCL on OpenCL backend}
\begin{figure}
    \centering
    \includegraphics[width=.7\linewidth]{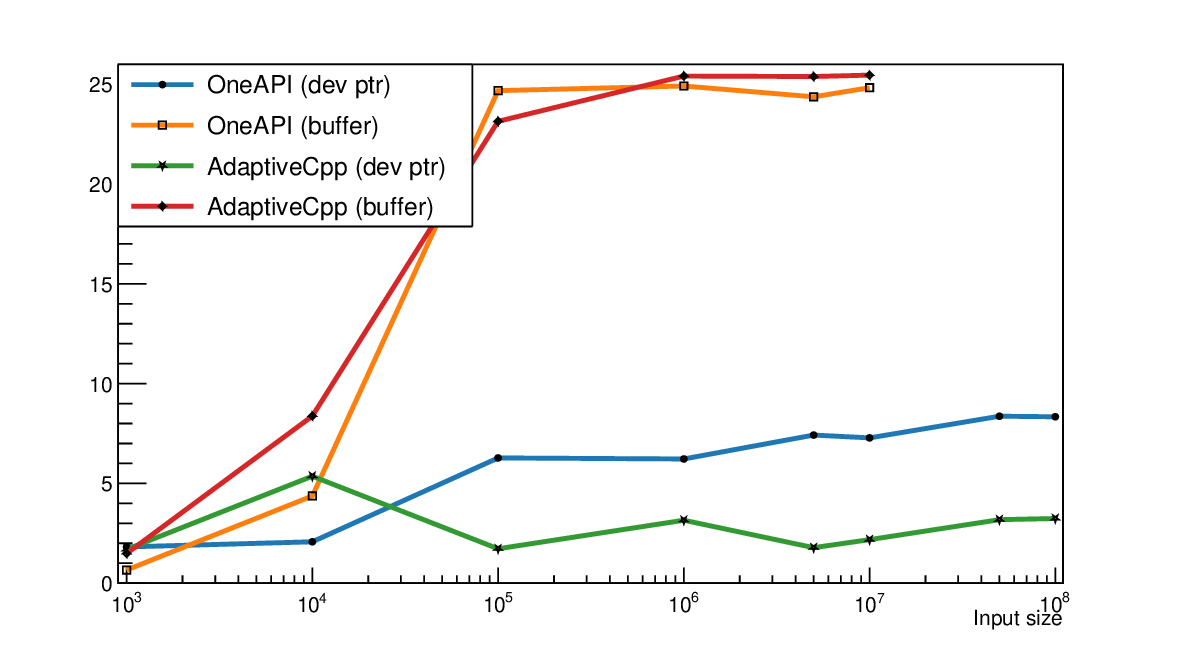}
    \includegraphics[width=.7\linewidth]{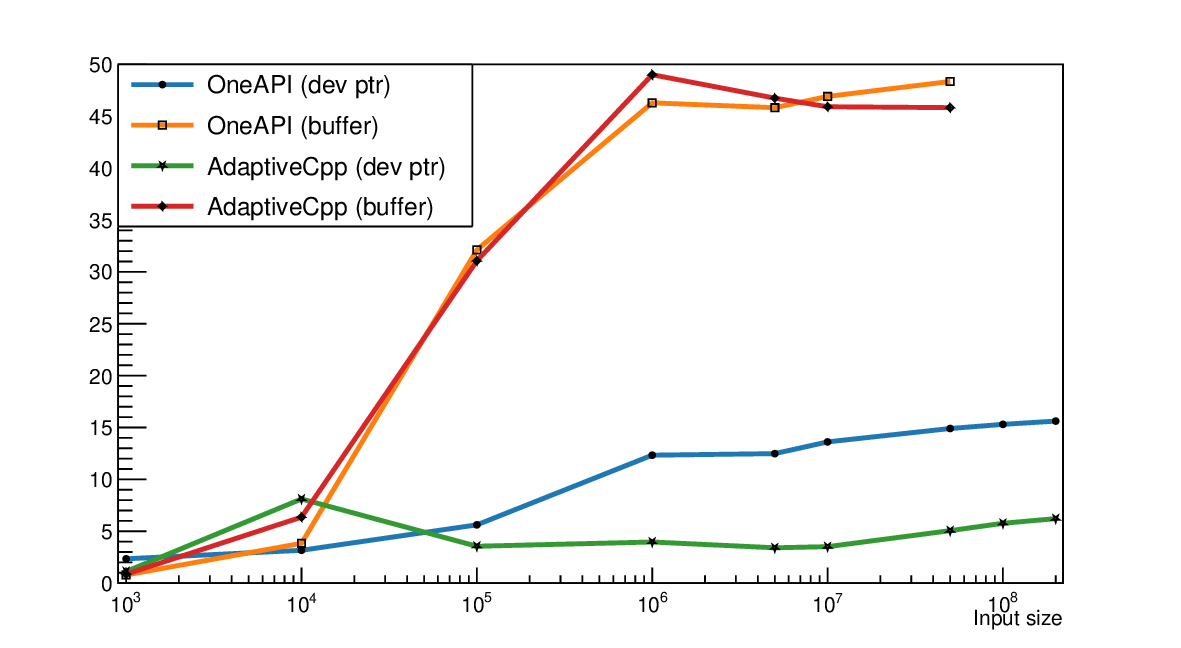}

    \caption{Weak scaling against single threaded execution on computing environment \ref{victus} for Boosting tests on OpenCL backend. Subfigure on the top show double precision results, while single precision results are shown on the bottom.}
    \label{fig:victus_ocl}
\end{figure}

We explore CPU performance by testing SYCL code on CPUs using OpenCL backend. Figure \ref{fig:victus_ocl} shows weak scaling results against single threaded execution on computing environment \ref{victus}. The subfigures only show results regarding the Boost test since results for the Invariant Masses test are comparable. Each subfigure shows the scaling obtained with Intel(R) OneAPI compiler with the use of device pointers and buffers, as well as the the scaling obtained with AdaptiveCpp compiler with the use of device pointers and buffers. Subfigure on the top show double precision results, while single precision results are shown on the bottom. 

\begin{figure}
    \centering
    \includegraphics[width=.7\linewidth]{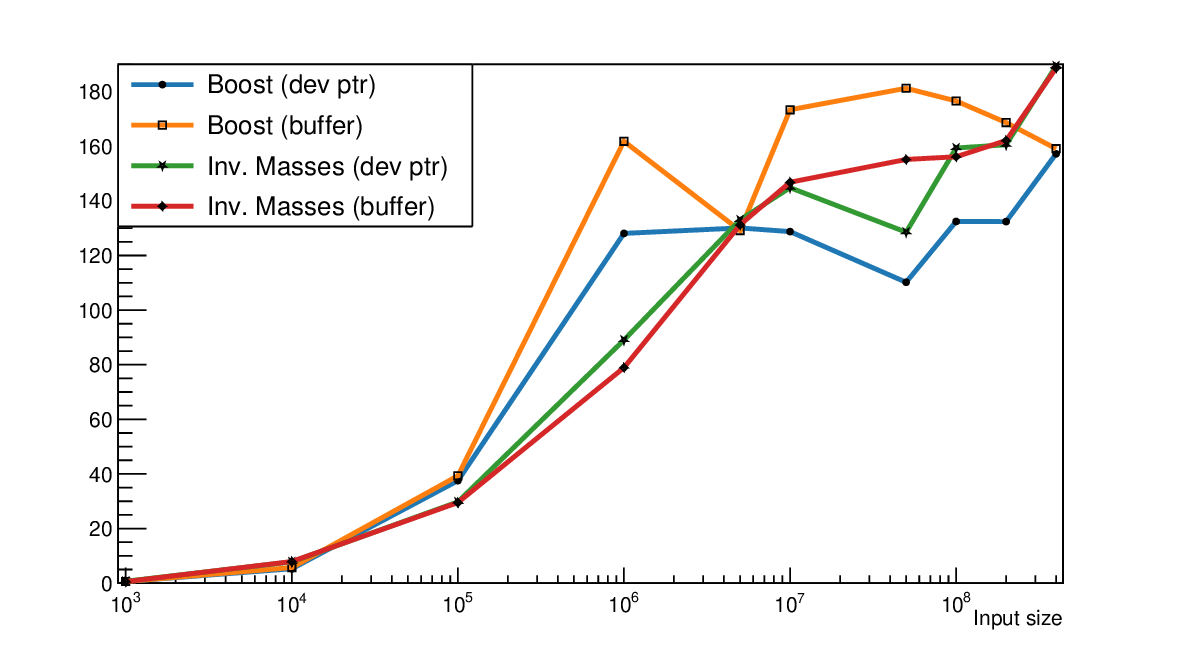}
    \includegraphics[width=.7\linewidth]{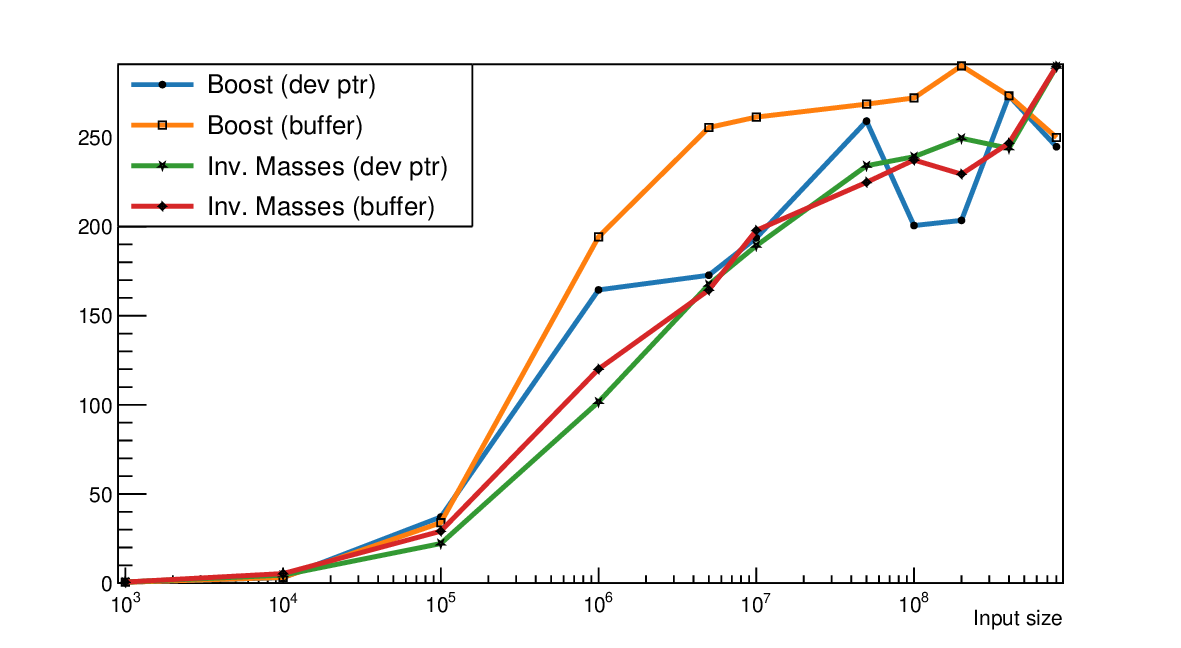}

    \caption{Weak scaling against single threaded execution on computing environment \ref{olgpu} for Boosting and Invariant Masses tests on OpenCL backend with Intel(R) OneAPI compiler. Subfigure on the top show double precision results, while single precision is shown on the bottom.}
    \label{fig:olgpu_ocl}
\end{figure}

\begin{figure}
    \centering
    \includegraphics[width=.7\linewidth]{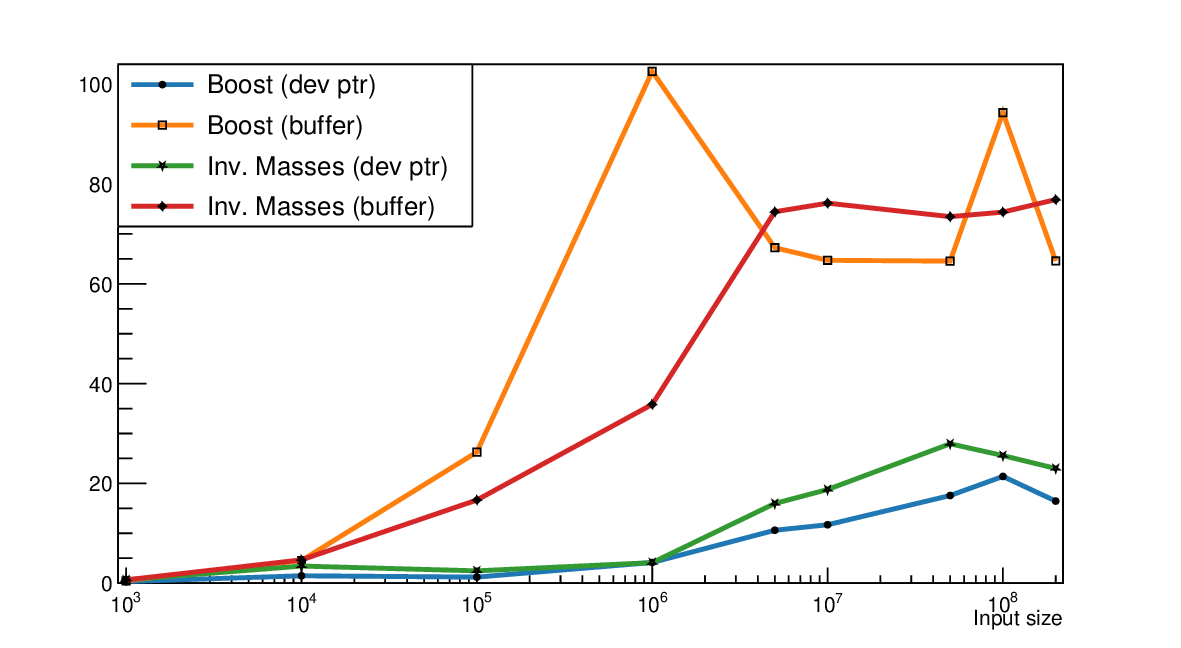}
    \includegraphics[width=.7\linewidth]{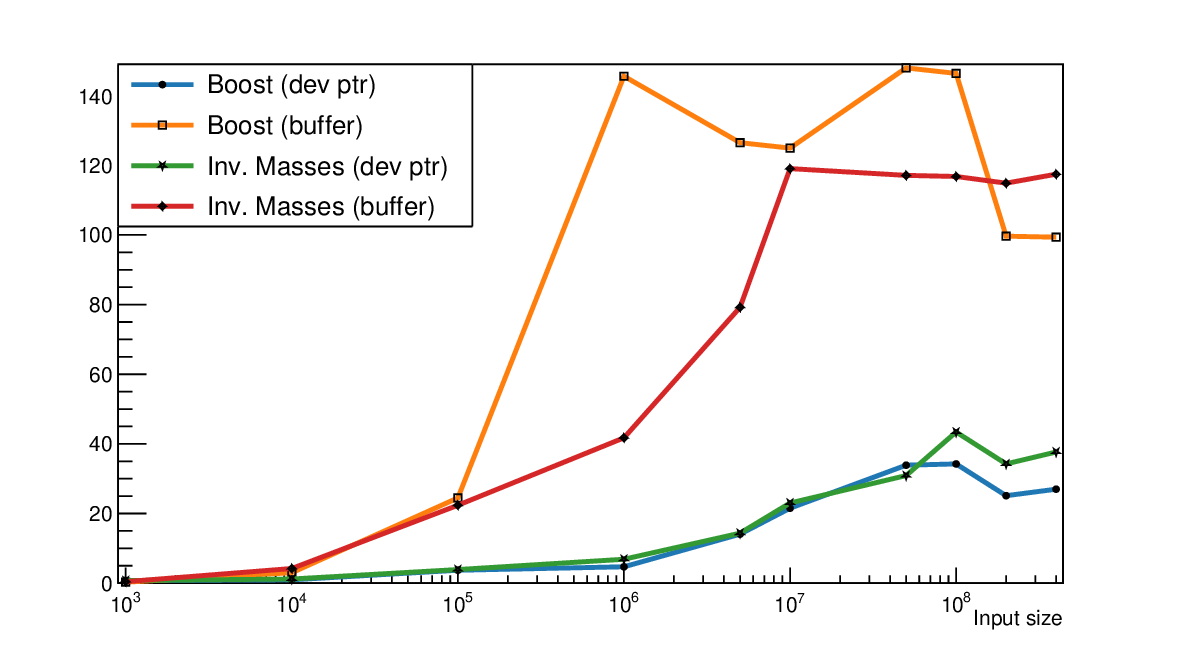}

    \caption{Weak scaling against single threaded execution on computing environment \ref{olice} for Boosting and Invariant Masses tests on OpenCL backend with Intel(R) OneAPI compiler. Subfigure on the top show double precision results, while single precision is shown on the bottom.}
    \label{fig:olice_ocl}
\end{figure}

Figures \ref{fig:olgpu_ocl} and \ref{fig:olice_ocl} show weak scaling results against single threaded execution on computing environments \ref{olgpu} and \ref{olice}, respectively. 
Each subfigure shows results for Boost and Invariant Masses tests for SYCL with Intel(R) OneAPI compiler with the use of device pointers and buffers. Subfigures on the left show double precision results, while single precision results are shown on the right. 

It is worth noticing that Figures \ref{fig:victus_ocl} and \ref{fig:olice_ocl} highlight a clear gap between the performance obtained with the use of buffers instead of device pointers, in an opposite way to what has been observed on CUDA backend. This can be easily explained by observing that device pointers introduce additional overhead on the CPU computation, but it is surprising to notice how significant it is.
The gap is less pronounced in Figure \ref{fig:olgpu_ocl} for what concerns double precision, but is it still recognizable for single precision results. Performance results concerning the Boost test with the use of buffers in Figures \ref{fig:olgpu_ocl} and \ref{fig:olice_ocl} show instabilities in the execution times whose root cause is hard to track down and out of the scope of the current work. 

\section{Conclusions and future work} \label{sec:conclusion}

In this work, we have described our efforts for creating GenVectorX, a SYCL and CUDA version of GenVector, a C++ package providing classes and functionality to particles involved in collisions in High Energy Physics research. This package is part of the ROOT library, a the tool for storing, analyzing and visualizing physics data regarding particle collisions adopted by physicists from all over the world. With this project, which is the first involving the SYCL framework in the HEP community to the best of our knowledge, we have explored the potentiality of SYCL as performance portable framework to migrate and modernize the fundamental GenVector package. We have carried out an extensive test campain, showing that SYCL can achieve a good performance portability with respect to CUDA and even outperform it. Furthermore, we have compared two main different implementations of SYCL compiler, namely AdaptiveCpp and Intel(R) OneAPI, showing that it is possible to achieve comparable performance with these tools. It is worth to remark that the use of two compiler has been particularly helpful in the context of code debugging. SYCL has shown impressive scaling on CPU backends as well, confirming its performance portability capacity with near-zero code divergence. These primary results encourage us to pursue the porting project to a production-ready level, which will next target the integration among the other packages in the ROOT library, in particular with user-familiar tools such as ROOTDataFrame (RDF) \cite{rdataframe}. 



\begin{acks}
This research was supported by the the Horizon Europe Project "SYCLOPS", funded from the European Union HE Research and Innovation programme under grant agreement No 10109287. This research used resources of the European Organization for Nuclear Research (CERN).
\end{acks}

\appendix

\section{Reproducibility of results}
We provide the code and detail the settings and parameters used in order to reproduce these results presented in this work.

\subsection{Obtaining the source code}
The source code is open-source and available on the \href{https://github.com/root-project/genvectorx}{ROOT project on Github}. 

\subsection{Building and Installing GenvectorX} 
To build GenVectoX, the CMake build platform is necessary., In this paper, CMake-3.22.1 was used on computing environment \ref{victus} and CMake-3.20.2 on computing environments \ref{olgpu} and \ref{olice}. To build GenVectoX with CUDA, the CUDA Toolkit needs to be installed. To enable SYCL support, one of the following installations is required:
\begin{itemize}
    \item Intel(R) oneAPI Toolkit, here version 2024.0.0 was used via installation script available \href{https://www.intel.com/content/www/us/en/developer/articles/tool/oneapi-standalone-components.html#dpcpp-cpp}{here}. 
    
    \item AdaptiveCPP, here it was built from source available on \href{https://github.com/AdaptiveCpp/AdaptiveCpp}{AdaptiveCpp project on Github} commit \\ \texttt{47b786430a5943a9649073ca2bee64c9980718bf}.
\end{itemize}

The installation used from results in this paper can be reproduced following these steps:
\begin{itemize}
    \item To build GenVectorX with Intel(R) OneAPI support: \\
    \texttt{\$ cmake .. -G Ninja -DCMAKE\_BUILD\_TYPE=Release \\
    -DSYCL\_ROOT\_DIR=/path/to/intel/oneapi/latest/ \\ 
    -Doneapi=ON -Dsyclcuda=ON \\
    -DDCMAKE\_CUDA\_ARCHITECTURES=8x \\
    -DCUDA\_TOOLKIT\_ROOT\_DIR=/path/to/cuda-12.x/ \\
    -DCMAKE\_INSTALL\_PREFIX=. \\
    -Dtesting=ON -Dsingle\_precision=ON}
    \item To build GenVectorX with AdaptiveCpp support: \\
    \texttt{\$ cmake .. -G Ninja -DCMAKE\_BUILD\_TYPE=Release \\ 
    -Dadaptivecpp=ON -DACPP\_TARGETS="omp;cuda:sm\_8x" \\
    -DAdaptiveCpp\_DIR=/path/to/AdaptiveCpp/lib/cmake\\/AdaptiveCpp/ \\ 
    -Doneapi=OFF -DCMAKE\_INSTALL\_PREFIX=. \\
    -Dtesting=ON -Dsingle\_precision=ON}
    \item To build GenVectorX with CUDA support: \\
    \texttt{\$ cmake .. -G Ninja -DCMAKE\_BUILD\_TYPE=Release \\
    -Dcuda=ON -DCMAKE\_CUDA\_HOST\_COMPILER=/path/to/g++ \\
    -DCUDA\_TOOLKIT\_ROOT\_DIR=/path/to/cuda-12.x \\
    -DCMAKE\_CUDA\_COMPILER=/path/to/cuda-12.x/bin/nvcc \\
    -DCMAKE\_CUDA\_ARCHITECTURES=8x \\
    -DCMAKE\_CUDA\_FLAGS="-arch=sm\_8x" \\
    -Dtesting=ON -Dsingle\_precision=ON \\
    -DCMAKE\_INSTALL\_PREFIX=.}
\end{itemize}
The installation can be finilized by running: \texttt{\$ ninja install}. All the steps above assume that a build directory has been created and accessed: \texttt{\$ mkdir build \&\& cd build}. The \texttt{testing} cmake variable enables the compilation of target tests, FP$64$ being the defaut precision. The \texttt{single\_precision} cmake variable enables the compilation enables computation of target tests in both single and double precision.
To build tests against ROOT library, a ROOT installation is required. In this paper, installation from source from \href{https://github.com/root-project/root}{ROOT project on Github} commit \texttt{f4a23f7e39c2a2b00e1c325db5d82206f70db0b6} was used. The installation used from results in this paper can be reproduced following these steps:  \\
\texttt{\$ cmake -DCMAKE\_INSTALL\_PREFIX=. \\
-DPython3\_EXECUTABLE=/path/to/python3 \\
-DCMAKE\_CXX\_STANDARD=17 \\
-DCMAKE\_CXX\_COMPILER=/path/to/g++ \\
-DCMAKE\_BUILD\_TYPE=Release \\
-Dtesting=ON -Droottest=ON ..} \\
The installation can be finilized by running: \texttt{\$ ninja install}. All the steps above assume that a build directory has been created and accessed: \texttt{\$ mkdir build \&\& cd build}.
\bibliographystyle{ACM-Reference-Format}
\bibliography{sample-base}

\end{document}